\documentclass[nofootinbib,floatfix,superscriptaddress,twocolumn]{revtex4}
\pdfoutput=1
\usepackage{graphicx}
\usepackage{epsfig}
\usepackage{bm}
\usepackage[T1]{fontenc}
\usepackage[latin9]{inputenc}
\usepackage{amssymb}
\usepackage{float}
\usepackage{amsmath}
\usepackage{dcolumn}
\usepackage{cancel}
\usepackage[colorlinks]{hyperref}
\usepackage[usenames,dvipsnames]{color}
\hypersetup{
     breaklinks=true,
    pdfstartview={FitH},    
    colorlinks=true,       
    linkcolor=blue,          
    citecolor=red,        
    filecolor=magenta,      
    urlcolor=blue,           
    anchorcolor=green,      
    linktocpage=true
}

\newcommand{\be}{\begin{equation}}
\newcommand{\ee}{\end{equation}}

\begin{document}

\title{Kaniadakis entropy-based characterization of IceCube PeV neutrino signals}

\author{M.~Blasone}
\email{blasone@sa.infn.it}
\affiliation{Dipartimento di Fisica, Universit\`a degli Studi di Salerno, Via Giovanni Paolo II, 132 I-84084 Fisciano (SA), Italy}
\affiliation{INFN Gruppo Collegato di Salerno - Sezione di Napoli c/o Dipartimento di Fisica, Universit\`a di Salerno, Italy}

\author{G.~Lambiase}
\email{lambiase@sa.infn.it}
\affiliation{Dipartimento di Fisica, Universit\`a degli Studi di Salerno, Via Giovanni Paolo II, 132 I-84084 Fisciano (SA), Italy}
\affiliation{INFN Gruppo Collegato di Salerno - Sezione di Napoli c/o Dipartimento di Fisica, Universit\`a di Salerno, Italy}

\author{G.~G.~Luciano}
\email{giuseppegaetano.luciano@udl.cat}
\affiliation{Applied Physics Section of Environmental Science Department, Escola Polit\`ecnica Superior, Universitat de Lleida, Av. Jaume
II, 69, 25001 Lleida, Spain}

\date{\today}

\begin{abstract}
Kaniadakis $\kappa$-thermostatistics is by now recognized as an effective paradigm to describe relativistic complex systems obeying power-law tailed distributions, as opposed to the classical (exponential-type) decay. It is founded on a non-extensive one-parameter generalization of the Bekenstein-Hawking entropy, which, in the cosmological framework, gives rise to modified Friedmann equations on the basis of the gravity-thermodynamic conjecture. Assuming the entropy associated with the apparent horizon of the Friedmann-Robertson-Walker
(FRW) Universe follows Kaniadakis prescription, in this work we analyze the observed discrepancy between the present bound on the Dark Matter relic abundance and the IceCube high-energy ($\sim 1\,\mathrm{PeV}$) neutrinos.
We show that this tension can be alleviated in the minimal model of Dark Matter decay with Kaniadakis-governed Universe evolution, while still considering the 4-dimensional Yukawa coupling between Standard Model and Dark Matter particles. This argument phenomenologically supports the need for a Kaniadakis-like generalization of the Boltzmann-Gibbs-Shannon entropy in the relativistic realm, opening new potential scenarios in high-energy astroparticle physics.
 \end{abstract}

 \maketitle

\section{Introduction}
\label{Intro} 

The IceCube Neutrino Observatory, a neutrino telescope within the glacial ice of the Geographic South Pole, extends over $1\,\mathrm{km}^3$ of ice from roughly $10^3\,\mathrm{m}$ under the surface~\cite{Ice0}. It was originally designed to search for neutrino sources in the TeV regime to explore the highest-energy astrophysical processes~\cite{Ice0,Ice1}.
Interestingly enough, some unexpected neutrino-initiated cascade events were also collected with PeV energies. While being initially attributed to astrophysical objects~\cite{Cholis,Anchordoqui}, these exotic signals were later understood to be likely unrelated to known hot-spots, like supernova remnants or active galactic nuclei~\cite{Aartsen}. Although gamma-ray bursts 
remain potential source candidates~\cite{GR1,GR2}, the most credited assumption is that these neutrinos may have been produced by the heavy decaying Dark Matter (DM)~\cite{DM1,DM2,DM3,DM4,DM5,DM6,DM8,DM9}. 

In~\cite{Chianese} Chianese and Merle 
speculated on the decay of a hypothetical thermal relic density of PeV scale DM via the minimal (4-dimensional) DM-neutrino Yukawa-like coupling $\mathcal{L}_{d=4}=y\hspace{0.3mm} \bar L \cdot H \chi$, where $\bar L$, $H$ and $\chi$ are the left-handed lepton doublet, Higgs doublet and the DM particle, respectively, while 
$y$ quantifies the interaction. Here, we have dropped for simplicity the index for the mass eigenstates of the three active neutrinos\footnote{For a recent discussion on whether to consider mass or flavor states as active part of neutrino interactions, see~\cite{FM1,FM2,FM3,FM3bis,FM4,FM5,FM6,FM7,FM8}.}. However, considerations on the optimal lifetime $\tau\sim10^{28}\,\mathrm{sec}$ of PeV DM~\cite{Unb1,Unb2} reveal that such a coupling fails to account for both the PeV DM relic abundance and the decay rate needed for IceCube~\cite{DM9,Boucenna,Chianese}. Though alternative mechanisms have been subsequently  invoked, including the existence of a secluded DM sector~\cite{Secluded}, freeze-out with resonantly enhanced annihilation~\cite{Dibari} or freeze-in~\cite{Fong,Roland,Fiorentin}, no definitive solution to the problem of IceCube PeV neutrinos has yet emerged. 

The DM model of~\cite{Chianese} is framed in the standard General Relativity (GR).  Nevertheless, empirical evidences from Type Ia Supernovae, CMB radiation and large-scale structures indicate that Einstein's theory and the ensuing cosmological Friedmann equations are to be properly corrected 
to comply with phenomenology and, in particular, to explain the late-time accelerating expansion of the Universe and the inflationary scenario~\cite{Vagnozzi2}. 

In the light of the gravity-thermodynamic conjecture~\cite{Jacob}, it is known that the Friedmann equations ruling the Universe evolution can be derived from the first law of thermodynamics on the apparent horizon of the Universe~\cite{FL1,FL2,FL3,FL5,FL6,FL7,FL8,FL10,FL11} along with the holographic principle and the Bekenstein-Hawking entropy~\cite{tHooft,Sussk}. Recently, arguments from different perspectives have converged on the idea that the conventional entropy-area law should be somehow generalized due to quantum gravitational~\cite{Rovelli,Ash,Adler,Zhang,Barrow,Calmet} and/or non-extensive~\cite{Iorio,TsallisCirto,Odin,Jizba:2022icu,LucBlas,SheNew} corrections. 

Among the most prominent examples of the latter class, Kaniadakis entropy arises from the effort to extend the classical Boltzmann-Gibbs statistics to the special relativistic context~\cite{Kania0,Kania1,Kania2,Scarf1}. In turn, the associated distribution computed through the maximum entropy principle is a one-parameter continuous deformation of the Maxwell-Boltzmann function, exhibiting power-law tails instead of the canonical exponential behavior.
Kaniadakis framework has so far been tested successfully for many high-energy systems, such as cosmic rays~\cite{Kania1}, plasma~\cite{plasma} and open stellar
clusters~\cite{OSC}. 

In parallel, one advantage of  
Kaniadakis entropy in Cosmology is the non-trivial impact it has on the predicted history of the Universe, which gets modified toward improving the $\Lambda$CDM model phenomenologically~\cite{Drepanou:2021jiv,CosmKan1,CosmKan2,CosmKan3,CosmKan4,CosmKan5,CosmKan6,Ghaf1} (see also~\cite{LucRev} for a recent review of Kaniadakis entropy applications in Gravity and Cosmology).
In particular, it is found that  the Hubble expansion rate acquires the form $H(T)=H_{GR}(T)Z_\kappa(T)$, where $H_{GR}$ is the rate obtained in the standard Cosmology based on GR, while $Z_\kappa$ contains Kaniadakis induced effects. Typically,
relativistic corrections are expected to be modulated over time, in such a way that
$Z_\kappa(T)\neq1$ in the earliest stages of the Universe existence (at the pre-BBN epoch, which is not directly constrained by cosmological observations), while it tends to unity at late-time, recovering classical GR. Although not considered in the original Kaniadakis framework, this behavior can be taken into account by assuming a running $\kappa$-parameter decreasing over time, as speculated in~\cite{CosmKan5}.   

Starting from the above premises, in this work we focus on the study of the observed discrepancy between the current bound on the Dark Matter relic abundance and the IceCube high-energy neutrino events in Kaniadakis Cosmology. In this respect, we would like to remark that alternative modified cosmologies based on deformed entropic scenarios have been proposed in~\cite{Tsallis1,Barrow1,Reny1,Moniz}  based on information theory or quantum gravitational considerations.
It is important to stress that the specific rationale behind the present analysis is that the PeV neutrinos revealed by IceCube are highly relativistic and, hence, more suited to be described in a picture that involves relativistic (Kaniadakis-like) statistical laws too.
In this context, we assert that the IceCube tension can be alleviated in a Universe governed by Kaniadakis-Cosmology implied Friedmann equations, while still employing the  minimal 4-dimensional interaction $\mathcal{L}_{d=4}$ defined above. We stress that this is a virtue of Kaniadakis formalism, which cannot be accounted for by the usual Cosmology. 

The remainder of the work is structured as follows: in the next Section we briefly review Kaniadakis statistics. Sec.~\ref{ModCosm} is devoted to discuss the modified Cosmology based on Kaniadakis horizon entropy, while in Sec.~\ref{IceCube} we apply Kaniadakis paradigm to the 
IceCube high-energy neutrino tension. Conclusions and outlook are finally summarized in Sec.~\ref{Disc}. Throughout the work, we use natural units $\hslash=c=k_B=1$, while we keep the gravitational constant $G$ explicit. In this way, we have $G=1/M_p^2$, with $M_p$ being the Planck mass.

\section{Kaniadakis statistics: a review}
\label{KanRev}

In this Section we discuss mathematical and physical basics of Kaniadakis statistics. For more details on the subject, see~\cite{Kania0,Kania1,Kania2}.  
It is known that the Maxwell-Boltzmann (MB) distribution is taken as a foundation of the classical statistical mechanics, rather than stemming from it. In fact, such a distribution 
emerges within the Newtonian mechanics, as suggested by numerical simulations of classical molecular dynamics~\cite{Ninarello}.

The question naturally arises as
to whether MB distribution is also
obtained within a framework governed by the special relativity 
laws at microscopic dynamical level. 
This problem has been addressed in~\cite{Kania0}, based on the evidence that the relativistic cosmic rays exhibit a power-law
tailed spectrum, in contrast to the MB  exponential behavior~\cite{Kania1}. The same feature has also been observed in other high-energy systems, such as 
the plasma in a superthermal radiation field~\cite{plasma}, nuclear collisions~\cite{NC} and open stellar clusters~\cite{OSC}. 
These evidences advise on the need to suitably generalize the classical Boltzmann-Gibbs-Shannon (BGS) entropic functional in the relativistic realm.

In~\cite{Kania1,Kania2} it has been shown the Lorentz transformations of the special relativity arguably 
impose the following  one-parameter deformation of Boltzmann-Gibbs entropy 
\begin{equation}
    \label{KE}
S_{\kappa}=-\sum_{i}n_i \ln_\kappa n_i\,,
\end{equation}
where the $\kappa$-logarithm is defined by
\begin{equation}
    \label{logk}
\ln_{\kappa}x \equiv \frac{x^\kappa-x^{-\kappa}}{2\kappa}\,.
\end{equation}

The generalized Boltzmann factor for the $i$-th level of the system
of energy $E_i$ takes the form
\begin{equation}
n_i=\alpha \exp_\kappa\left[-\beta\left(E_i-\mu\right)\right],
\end{equation}
where 
\begin{equation}
\label{expk}
\exp_\kappa(x)\,\equiv\,\left(\sqrt{1+\kappa^2\,x^2}\,+\,\kappa\,x\right)^{1/\kappa}\,,
\end{equation}
and
\begin{eqnarray}
\alpha&=&\left[(1-\kappa)/(1+\kappa)\right]^{1/2\kappa}\,,\\[2mm] 1/\beta&=&\sqrt{1-\kappa^2}\,\hspace{0.2mm}T\,.
\end{eqnarray}

The deformed entropy in Eq.~\eqref{KE} is known as \emph{Kaniadakis entropy}. Deviations from the classical framework are quantified by the dimensionless parameter $-1<\kappa<1$, in such a way that the standard statistics is recovered  in the Galilean $\kappa\to 0$ limit. For later convenience, we remind that Kaniadakis entropy can be equivalently expressed  as~\cite{Jeans1,Jeans2}
\be
\label{KanP}
S_\kappa=-\sum_{i=1}^W \frac{P_i^{1+\kappa}-P_i^{1-\kappa}}{2\kappa}\,,
\ee
where $P_i$ denotes the probability of the system to be in the $i$-th microstate and $W$ is the total number of configurations.

To have more contact with the physical language, it is now convenient to introduce the following functions
\begin{eqnarray}
\label{u}
u(q)&=&\frac{q}{\sqrt{1+\kappa^2q^2}}\,,\\[2mm]
\mathcal{W}(q)&=&\frac{1}{\kappa^2}\sqrt{1+\kappa^2q^2}-\frac{1}{\kappa^2}\,,\\[2mm]
\varepsilon(\mathcal{W})&=&\mathcal{W}+\frac{1}{\kappa^2}\,,
\label{eps}
\end{eqnarray}
which correspond to the (auxiliary) dimensionless velocity, kinetic  and total energy of a given one-particle system, respectively. Here, we have denoted the dimensionless momentum by $q$.

The above relations can be easily inverted to give
\begin{eqnarray}
q(u)&=&\frac{u}{\sqrt{1-\kappa^2u^2}}\,,\\[2mm]
\mathcal{W}(\varepsilon)&=&\varepsilon-\frac{1}{\kappa^2}\,,\\[2mm]
\varepsilon(q)&=&\frac{1}{\kappa^2}\sqrt{1+\kappa^2q^2}\,.
\end{eqnarray}

At this stage we can define the physical velocity $v$, momentum $p$ and total energy $E$ through~\cite{Kania2} 
\begin{equation}
\label{qdef}
    \frac{v}{u}=\frac{p}{mq}=\sqrt{\frac{E}{m\epsilon}}=\kappa c \equiv v_*
    \,.
\end{equation}
Similarly, the kinetic energy is given by
\begin{equation}
    W=E-mc^2\,,
\end{equation}
with $m$ being the rest mass of the system.
In order for these variables to be consistently defined in the Galilean limit too, we have to require 
\begin{equation}
 \lim_{c\rightarrow\infty,\kappa\rightarrow0} v_*<\infty\,.   
\end{equation}
In so doing, insertion of the physical variables into Eqs.~\eqref{u}-\eqref{eps} allows us to recover the standard momentum/energy formulas of a particle in the special-relativistic regime, i.e.
\be
p=\gamma m v\,,\,\,\,\,\,\,\, E=\gamma mc^2\,,
\ee
where $\gamma=1/\sqrt{1-v^2/c^2}$ is the relativistic Lorentzian factor. 

A comment is in order here: besides Kaniadakis formulation, 
other relativistic generalizations of the MB distribution have been proposed in the literature. Among these, Maxwell-J\"{u}ttner velocity distribution~\cite{MJ} represents the first attempt toward the construction of a relativistic statistical theory. Such a model, however, is developed by naively replacing the relativistic energy-velocity relation into the Maxwell-Boltzmann factor. In turn, this gives rise to a hybrid distribution, which still maximizes the classical BGS entropy. On the other hand, Kaniadakis distribution~\eqref{expk} is derived ab initio from an entropic functional compatible with the special relativity, thus providing a self-consistent relativistic statistical framework.

\section{Modified Cosmology through
Kaniadakis horizon entropy}
\label{ModCosm}

Let us now export Kaniadakis paradigm to the black-hole framework. This step will then be
useful for the holographic (and, hence, cosmological) application of Kaniadakis entropy. Toward this end, we assume equiprobable states $P_i=1/W$ in Eq.~\eqref{KanP} and use the property that the Boltzmann-Gibbs-Shannon entropy is $S\propto\log W$. Since the Bekenstein-Hawking entropy is $S_{BH}=A/(4G)$, we have $W=\exp\left[A/(4G)\right]$. 
By plugging into Eq.~\eqref{KanP}, we find
\be
\label{KenBH}
S_\kappa=\frac{1}{\kappa}\sinh\left(\kappa\hspace{0.4mm}\frac{A}{4G}\right),
\ee
which indeed recovers the standard 
Bekenstein-Hawking entropy $S_{BH}$ for $\kappa\rightarrow0$. 

Notice that, since the above function is even,  i.e. $S_\kappa=S_{-\kappa}$, we can 
safely restrict to the $\kappa>0$ domain for our next considerations. In addition, given that deviations from the Bekenstein-Hawking formula are expected to be small, 
it is reasonable to approximate Eq.~\eqref{KenBH} for $\kappa\ll1$ as
\be
S_{\kappa}=S_{BH}+\frac{\kappa^2}{6}S_{BH}^3+\mathcal{O}(\kappa^4)\,,
\ee
where the first term is the usual entropy, while the second one provides the leading-order 
Kaniadakis correction.

We can now proceed with the derivation of
the $\kappa$-modified Friedmann equations. For this purpose, we follow~\cite{CosmKan2} and describe the 4-dim. background by a homogeneous and isotropic (Friedmann-Robertson-Walker) flat geometry with metric 
\be
\label{FRW}
ds^2=-dt^2+a^2(t)\left(dr^2+r^2d\Omega^2\right),
\ee
where $a(t)$ denotes the time-dependent scale factor and  ${\rm d}\Omega^2={\rm d}\theta^2+\sin^2\theta {\rm d}\varphi^2$ 
is the angular part of the
metric on the two sphere.
Moreover, 
we assume that the Universe is filled with a matter perfect fluid of mass density $\rho_m$ and pressure $p_m=w\rho_m$ at equilibrium, where $-1\leq w\leq1/3$ is the equation-of-state parameter. 

As a next step, we apply the gravitational thermodynamics conjecture to the Universe apparent horizon of radius $r_a=1/H=a/\dot a$ and effective temperature $T=1/(2\pi r_a)$. Practically, this consists in using the first law of thermodynamics
\be
\label{therm}
dU=TdS-WdV\,,
\ee
on the horizon of the Universe, which is conceived
as a (spherical) expanding thermodynamic system. Here, $W=(\rho_m-p_m)/2$ is the work density due to the change in the apparent horizon radius of the Universe, while $dU$ and $dV$ are the corresponding increase in internal energy and volume, respectively. Observing that $dU=-dE$, where $E=\rho_m V$ is the total energy content inside the Universe of volume $V=4\pi r_a^3/3$, Eq.~\eqref{therm} can be equivalently cast as
\be
\label{therm2}
dE=-TdS+WdV\,.
\ee

We now follow~\cite{Weinberg}, but with the generalized Kaniadakis entropy~\eqref{KenBH} instead of the Bekenstein-Hawking one. Omitting standard textbook calculations, we get from Eq.~\eqref{therm2}~\cite{CosmKan2}
\be
\label{FFE}
-4\pi G\left(\rho_m+p_m\right)=\cosh\left(\kappa\frac{\pi}{G H^2}\right)\dot H\,,
\ee
where the overdot indicates derivative respect to the cosmic time $t$. Furthermore, by imposing the conservation equation
\be
\nabla_{\mu}T^{\mu\nu}=0\,,
\ee
for the matter fluid of stress-energy tensor 
\be
T_{\mu\nu}=(\rho_m+p_m)u_\mu u_{\nu}+p_m\,g_{\mu\nu}\,,
\ee
and  four-velocity $u_\mu$, we are led to
\be
\label{ce}
\dot \rho_m=-3H(\rho_m+p_m)\,.
\ee
After substitution into Eq.~\eqref{FFE}, integration of both sides gives~\cite{CosmKan2}
\be
\label{SFE}
\frac{8\pi G}{3}\rho_m=\cosh\left(\kappa\frac{\pi}{G H^2}\right)H^2-\frac{\kappa\pi}{G}\mathrm{shi}\left(\kappa\frac{\pi}{GH^2}\right),
\ee
where we have set the integration (i.e. cosmological) constant to zero and we have defined
\be
\mathrm{shi}(x)\equiv \int_{0}^{x} \frac{\sinh(x')}{x'}\, dx'\,.
\ee

The relations~\eqref{FFE} and~\eqref{SFE} are the modified Friedmann equations underlying Kaniadakis Cosmology. They represent the central ingredient for the investigation of the evolution of the Universe. We emphasize that the extra $\kappa$-dependent corrections give rise to fascinating cosmic scenarios with a richer phenomenology comparing to the standard $\Lambda$CDM model. For instance, in~\cite{Drepanou:2021jiv} a holographic dark energy description based on Eqs.~\eqref{FFE} and~\eqref{SFE} has served to explain the current accelerated expansion of the Universe, while 
in~\cite{CosmKan5} the baryogenesis and primordial Lithium abundance problems have been successfully addressed.  
It is easy to check that the General Relativity framework is correctly recovered in the Bekenstein-Hawking entropy $\kappa\rightarrow0$ limit. 

In passing, we mention  
that modified Friedmann equations in alternative entropic scenarios have also been studied in Tsallis~\cite{Tsallis1,Tsallis2,Tsallis3,Tsallis4,Tsallis5} and Barrow~\cite{Barrow1,Barrow2,Barrow3,Barrow5, Barrow6,Barrow7,Barrow8} Cosmologies, motivated by non-extensive and quantum gravitational considerations, respectively~\cite{Vagnozzi}. Along this line, the IceCube PeV neutrino discrepancy has been 
examined in~\cite{JizLamb} in Tsallis Cosmology to constrain the related entropic parameter. In this sense, our next analysis 
resembles that of~\cite{JizLamb} and, more general, of~\cite{Stabile,CapozLamb} in extended theories of gravity. Here, however, we stress that corrections brought about in the Friedmann equations arise from a genuinely relativistic  deformation of the entropy-area law, rather than a modification of the gravitational interaction. 

\section{High-energy neutrino signals from IceCube}
\label{IceCube}
In this Section we 
present the useful 
features related to DM
relic abundance and IceCube data. To describe the interaction between  Standard
Model and Dark Matter particles,  we use the minimal (4-dimensional) Yukawa-like coupling 
\be
\label{Yc}
\mathcal{L}_{d=4}=y_{\sigma\chi}\hspace{0.3mm} \bar L_\sigma \cdot H \chi\,,
\ee
where $\sigma=e,\mu,\tau$ labels the eigenstates of the
three active neutrinos, $L_\sigma$ and $H$ are the left-handed lepton 
and Higgs doublets, respectively, $\chi$ the DM particle and $y_{\sigma\chi}$ the (dimensionless)
Yukawa coupling constants. Computations are first developed in the conventional Cosmology, showing that it is unable to reconcile the current bound on DM relic abundance
and IceCube high-energy events of neutrinos. We then argue that this controversy can be avoided, provided that the background evolution is described by Kaniadakis entropy-based Cosmology.

\subsection{Standard Cosmology}
\label{SCsec}
Following~\cite{JizLamb,Dhuria,Eise}, we consider the so called DM \emph{freeze-in} production, which means that DM particles
are never in thermal equilibrium
due to their weak interactions and are produced from the hot
thermal bath. If we define the DM abundance by $Y_\chi=n_\chi/s$, where 
$n_\chi$ is the number density
of DM particles, $s=2\pi^2g_*(T)T^3/45$ the entropy density and $g_*(T)\simeq106.75$ the effective number of degrees of freedom, the evolution equation for DM particles in the traditional Cosmology reads~\cite{Dhuria}
\be
\frac{dY_\chi}{dT}=-\frac{1}{H_{GR}(T)Ts}\frac{g_\chi}{\left(2\pi\right)^3}\int C \hspace{0.2mm}\frac{d^3p_\chi}{E_\chi}\,,
\ee
where $H_{GR}$ is the standard Hubble
rate of General Relativity, $g_\chi=2$ the two helicity projections of DM and $C$ the general
collision term. The momentum and energy scale of DM have been denoted by $p_\chi$ and $E_\chi$, respectively. 

For constant $g_*$, the DM
relic abundance can be written as~\cite{Dhuria}
\be
\label{ODM}
\Omega_{DM}h^2=\left|\frac{2m_\chi^2s_0h^2}{\rho_{c}}\int_0^\infty\frac{dx}{x^2}\left(-\frac{dY_\chi}{dT}\Big|_{T=\frac{m_\chi}{x}}
\right)
\right|\,,
\ee
where $x\equiv m_\chi/T$, $m_\chi$ is the DM mass scale and $h$ the dimensionless Hubble constant. 
Furthermore, the present value of the entropy density and the critical
density have been indicated by
\begin{eqnarray}
\label{s0}
s_0&=&2\pi^2g_*T_0^3/45\simeq2891.2/\mathrm{cm^3}\,,\\[2mm]
\rho_c&=&1.054\times 10^{-5} h^2\,\mathrm{GeV/cm^3}\,,
\label{rhoc}
\end{eqnarray}
respectively. 

For the observed DM abundance, Eq.~\eqref{ODM} gives the value~\cite{ODMA}
\be
\label{obs}
\Omega_{DM}h^2\big|_{obs}=0.1188\pm0.0010\,.
\ee
Now, the most relevant processes that are induced by the interaction~\eqref{Yc} and contribute
to the DM production are the \emph{inverse decays}
\be
\label{inv}
\emph{i})\,\,\,\,\,\, \nu_\sigma+H^0\rightarrow\chi\,, \quad\, l_\sigma+H^+\rightarrow\chi\,,
\ee
and the \emph{Yukawa production} processes 
\be
\emph{ii})\,\,\,t+\bar t\rightarrow \bar\nu_\sigma+\chi\,.
\ee
While the former are kinematically allowed, provided that $m_\chi>m_H+m_{\nu,l}$ and have probabilities proportional to $|y_{\sigma\chi}|^2$, the latter  
depend on the factor $|y_{\sigma\chi}y_t|^2$, where 
$t$ is the top quark and $y_t$ the Yukawa coupling 
constant between the top quark and Higgs
boson. Thus, the evolution of DM
particles induced by the interaction~\eqref{Yc} 
becomes~\cite{Dhuria}
\be
\frac{dY_\chi}{dT}=\frac{dY_\chi}{dT}\Big|_{\emph{i)}}+\frac{dY_\chi}{dT}\Big|_{\emph{ii)}}\,,
\ee
where 
\begin{eqnarray}
\label{inversedec}
\frac{dY_\chi}{dT}\Big|_{\emph{i)}}&=&-\frac{m_\chi^2\Gamma_\chi}{\pi^2 H_{GR}(T) s}K_1\left(\frac{m_\chi}{T}\right)\,,\\[2mm]
\nonumber
\frac{dY_\chi}{dT}\Big|_{\emph{ii)}}&=&-\frac{1}{512\pi^6 H_{GR}(T) s}\int d\bar s d\Omega\\[2mm]
&&\sum_\sigma\frac{W_{t\bar t\rightarrow\bar\nu_\sigma\chi}+2W_{t\nu_\sigma\rightarrow t\chi}}{\sqrt{\bar s}}K_1\left(\frac{\sqrt{\bar s}}{T}\right).
\end{eqnarray}
Here, $\bar s$ represents the centre-of-mass energy,
$W_{ij\rightarrow kl}$ are the scattering probabilities of the related processes, $K_1(x)$ is the modified Bessel function of the second
kind and 
\be
\label{intrate}
\Gamma_\chi=\sum_\sigma \frac{|y_{\sigma\chi}|^2}{8\pi}m_\chi
\ee
the interaction rate. 

As argued in~\cite{Dhuria}, 
the very dominant processes in the DM production are the inverse decays~\eqref{inv}. Accordingly, the DM relic abundance is approximately
\be
\Omega_{DM}h^2\big|_{\emph{i)}}\simeq0.1188\frac{\sum_\sigma |y_{\sigma\chi}|^2}{7.5\times10^{-25}}\,.
\ee

Therefore, the observed value~\eqref{obs} is reproduced, provided that $\sum_\sigma |y_{\sigma\chi}|^2\simeq7.5\times10^{-25}$. This is, however, at odds with the condition 
required to fit the IceCube high-energy neutrino events. Indeed, let us notice that the stability of DM particles imposes that the lifetime $\tau_\chi=\Gamma_\chi^{-1}$ 
has to be longer than the age of the Universe, i.e. $\tau_\chi>t_U\simeq4.35\times10^{17}\,\mathrm{sec}$. Furthermore, the IceCube spectrum sets the (nearly model-independent) more stringent lower bound $\tau_\chi\gtrsim\tau_\chi^b\simeq10^{28}\,\mathrm{sec}$~\cite{Chianese}. By plugging the aforementioned estimate $\sum_\sigma |y_{\sigma\chi}|^2\simeq7.5\times10^{-25}$ into Eq.~\eqref{intrate}, one obtains $\Gamma_\chi\simeq 4.5\times 10^4\frac{m_\chi}{\mathrm{PeV}}\mathrm{sec^{-1}}$, which in turn implies $\tau_\chi\simeq2.2\times 10^{-5}\frac{\mathrm{PeV}}{m_\chi}\mathrm{sec}$. For $m_\chi\simeq1\mathrm{PeV}$, we then have $\tau_\chi\simeq2.2\times10^{-5}\,\mathrm{sec}$, in contrast with what stated above. 

On the other hand, in order to be compatible with the DM decay lifetime $\tau_\chi\simeq10^{28}\,\mathrm{sec}$ required by IceCube, we should have 
\be
\label{ICdata}
\sum_\sigma |y_{\sigma\chi}|^2_{IceCube}\simeq1.6 \times 10^{-57}\,,
\ee
which is 
by far (roughly 33 orders of magnitude) lower  than the value needed to explain the DM relic abundance. 

The above considerations make it clear that the IceCube high energy events and the DM relic abundance are inconsistent with the DM production as far as the latter is ascribed to the interaction~\eqref{Yc} and the cosmological background evolves according to the Einstein field equations.

\subsection{Kaniadakis entropy-based Cosmology}

Let us now explore how the above picture is modified in Kaniadakis Cosmology. To extract analytical solution, it proves convenient to perform Taylor expansion of the Friedmann equation~\eqref{SFE} for small $\kappa$, which is indeed the case according to  the discussion below Eq.~\eqref{KenBH}.  Observing that 
\begin{eqnarray}
\cosh(x)&=&1+\frac{x^2}{2}+\frac{x^4}{24}+\mathcal{O}(x^6)\,,\\[2mm] 
\mathrm{shi}(x)&=&x+\frac{x^3}{18}+\frac{x^5}{600}+\mathcal{O}(x^7)\,,
\end{eqnarray}
we get to the leading order
\be
\frac{8\pi G}{3}\rho_m\simeq H^2-\kappa^2\hspace{0.2mm}\frac{\pi^2}{2\left(GH\right)^2}\,.
\ee
This equation can be solved with respect to $H$
to obtain
\begin{eqnarray}
\nonumber
H&\simeq&\left[\frac{4\pi G \rho_m}{3}+\frac{\pi\left(64G^6\rho_m^2+18G^2\kappa^2\right)^{\frac{1}{2}}}{6G^2}\right]^{\frac{1}{2}}\\[2mm]
&\simeq&H_{GR}+\sqrt{\frac{27\pi}{2}}\frac{\kappa^2}{64\left(G^7\rho_m^3\right)^{\frac{1}{2}}}\,,
\label{Hmod}
\end{eqnarray}
where we have only considered the solution that, for $\kappa\rightarrow0$, recovers the correct limit 
\be
\label{Hst}
H_{GR}=\sqrt{\frac{8\pi G}{3}\rho_m}\,.
\ee 

As explained in Sec.~\ref{Intro}, in order to isolate corrections arising from modified gravity, it is useful to factorize the Hubble rate~\eqref{Hmod} as
\be
\label{MHR}
H(T)=H_{GR}(T) Z_\kappa(T)\,,
\ee
where the information on the modified Kaniadakis entropy is contained in the extra factor
\be 
\label{Zk}
Z_\kappa(T)\simeq 1+ \frac{9\kappa^2}{256\left(G^2\rho_m\right)^2}\,.
\ee
Some comments are in order: first, we notice that
the $\kappa\rightarrow0$ limit of Eq.~\eqref{Zk} gives $Z_\kappa=1$, as expected. Though being derived
in a different way, Eq.~\eqref{Zk} is consistent with the result of~\cite{CosmKan5}. Moreover, we can relate the matter density and temperature as
\be
\rho_m=\frac{\pi^2g_*(T)}{30}T^4\,,
\ee
where $g_*(T)\simeq106.75$ as defined in the previous Section. 

The usage of the modified Hubble rate~\eqref{MHR}
allows us to recast the evolution equation~\eqref{inversedec} of DM particles produced by the inverse decays as
\be
\label{Abund}
\frac{dY_\chi}{dT}\Big|_{\emph{i)}}=-\frac{m_\chi^2\Gamma_\chi}{\pi^2 H(T) s}K_1\left(\frac{m_\chi}{T}\right)\,,\\[2mm]
\ee
where now
\be
\label{sH}
H(T)s\simeq \frac{64\pi^4g_*^2\hspace{0.3mm} T^8+2025\hspace{0.3mm}T_*^8\kappa^2}{2160\hspace{0.3mm}\sqrt{5\pi g_*}\hspace{0.3mm}T^3\hspace{0.3mm} T_*}\,,\quad T_*=M_p=\frac{1}{\sqrt{G}}\,,
\ee
to the leading order in $\kappa$.  Here, $s$ is the entropy density defined at the beginning of Sec.~\ref{SCsec}. 

Employing Eqs.~\eqref{Abund} and~\eqref{sH} and following the same computations as in Sec.~\ref{SCsec}, 
the $\kappa$-modified DM relic abundance~\eqref{ODM} becomes 
\begin{eqnarray}
\Omega_{DM}h^2&=&\left|\frac{2m_\chi^2s_0h^2}{\rho_{c}}\int_0^\infty\frac{dx}{x^2}\left(-\frac{dY_\chi}{dT}\Big|_{T=\frac{m_\chi}{x}}
\right)
\right|\\[2mm]
\nonumber
&&\hspace{-19mm}\simeq\frac{3.5\hspace{0.5mm}h^2s_0\hspace{0.3mm}\Gamma_\chi\hspace{0.3mm}T_*}{\pi^{\frac{17}{2}}\hspace{0.1mm}g_*^{\frac{7}{2}}\hspace{0.3mm}m_\chi^9\hspace{0.3mm}\rho_c}\left|64\pi^4g_*^2m_\chi^8 -6.6\times10^{9}\hspace{0.3mm}T_*^8\kappa^2\right|\,,
\end{eqnarray}
where we have used~\cite{AeS}
\be
\int_0^{\infty}x^n K_1(x)\,dx=2^{n-1}\Gamma\left(1+\frac{n}{2}\right)\Gamma\left(\frac{n}{2}\right),\quad \Re[n]>0\,.
\ee

By further substituting Eq.~\eqref{intrate}, we get
\begin{eqnarray}
\nonumber
\Omega_{DM}h^2&\simeq&\frac{0.4\hspace{0.5mm}h^2s_0\hspace{0.3mm}T_*}{\pi^{\frac{19}{2}}\hspace{0.1mm}g_*^{\frac{7}{2}}\hspace{0.3mm}m_\chi^8\hspace{0.3mm}\rho_c}\sum_{\sigma}|y_{\sigma\chi}|^2\\[2mm]
&&\times \left|64\pi^4g_*^2m_\chi^8 -6.6\times10^{9}\hspace{0.3mm}T_*^8\kappa^2\right|\,.
\end{eqnarray}

For comparison with observational data, it is useful to cast the above expression as
\be
\label{recast}
\Omega_{DM}h^2\simeq0.1188\left(\frac{106.75}{g_*}\right)^{\frac{3}{2}}\frac{\sum_{\sigma}|y_{\sigma\chi}|^2}{1.6\times10^{-57}}\,\Pi_\kappa\,,
\ee
where we have defined
\begin{eqnarray}
\nonumber
\Pi_\kappa&\simeq&6.3\times10^{-61}\hspace{0.3mm}\frac{h^2s_0T_*}{\rho_c}\left|1-\frac{10^6\hspace{0.3mm}T_*^8\hspace{0.3mm}\kappa^2}{g_*^2\hspace{0.3mm}m_\chi^8}\right|\\[2mm]
&\simeq&1.7\times10^{-52}\frac{T_*}{1\,\mathrm{GeV}}\left|1-\frac{10^6\hspace{0.3mm}T_*^8\hspace{0.3mm}\kappa^2}{g_*^2\hspace{0.3mm}m_\chi^8}\right|\,.
\label{Pi}
\end{eqnarray}
In the second step we have used 
Eqs.~\eqref{s0} and~\eqref{rhoc} for $s_0$ and $\rho_c$, respectively. 

From Eq.~\eqref{recast} and~\eqref{Pi}, it follows that the DM relic abundance~\eqref{obs} and the IceCube data~\eqref{ICdata} are successfully and simultaneously 
explained in Kaniadakis Cosmology, provided that 
\be
\label{cond}
\Pi_\kappa\simeq1\,.
\ee

\begin{figure}[t]
\begin{center}
\hspace{-1mm}\includegraphics[width=8.7cm]{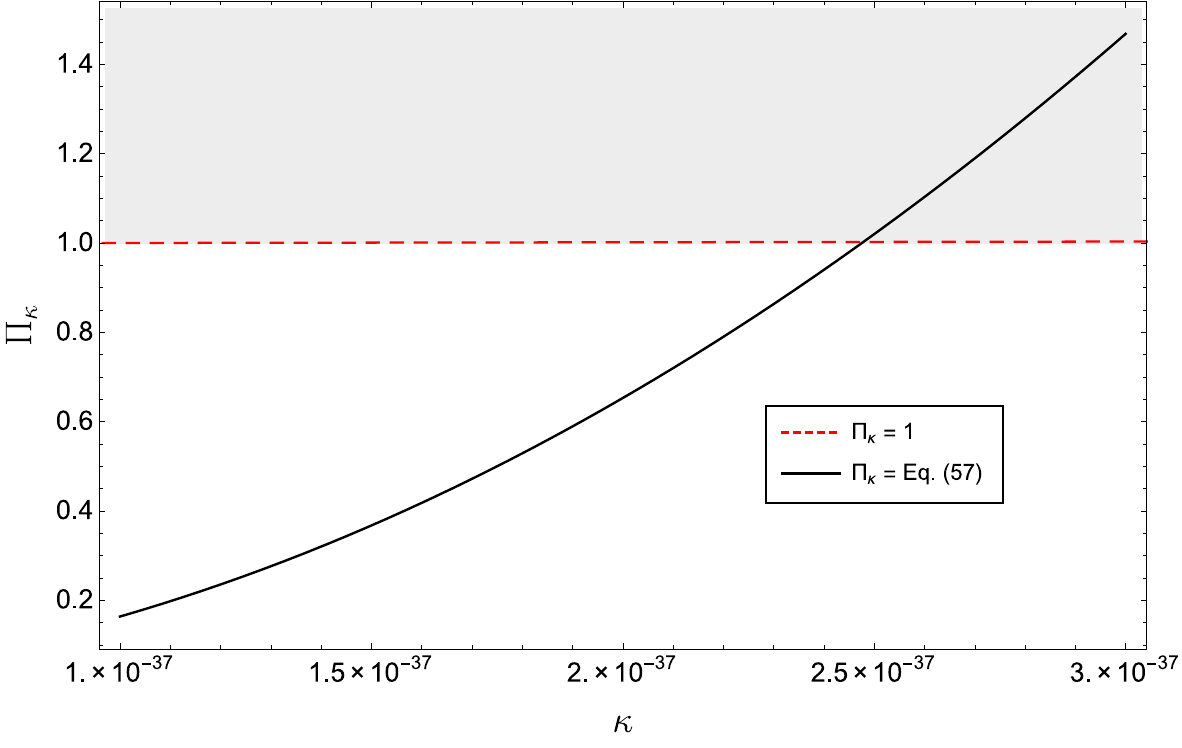}
\caption{\it{Plot of $\Pi_\kappa$ in Eq.~\eqref{Pi} versus $\kappa$ (black solid line). We have set the DM mass $m_\chi\simeq 1\,\mathrm{PeV}$ and $T_*\simeq 10^{19}\,\mathrm{GeV}$. The shaded region is phenomenologically forbidden, while the observed DM  abundance $\Omega_{DM}h^2\simeq0.1188$ in Eq.~\eqref{obs} is obtained for $\Pi_\kappa\simeq1$ (red dashed line).}}
\label{Fig1}
\end{center}
\end{figure}

The behavior of Eq.~\eqref{Pi} versus the Kaniadakis parameter $\kappa$ is plotted in Fig.~\ref{Fig1} for $m_\chi\simeq 1\,\mathrm{PeV}=10^6\,\mathrm{GeV}$ and the energy scale $T_*=M_p\simeq 10^{19}\,\mathrm{GeV}$. We observe that the condition~\eqref{cond} is satisfied, provided that
\be
\label{estimate}
\kappa\simeq2.5\times10^{-37}\,,
\ee 
which substantiates a posteriori our working assumption $\kappa\ll1$. 
It should be noted that, for the considered values of $m_\chi$ and $T_*$, a resolution of the problem going beyond the leading order approximation would be advisable. 
This, however, does not undermine the
conceptual validity of our assertion, that is the need for a relativistic generalization of the statistical framework (and, in particular, of the entropy-area law) to explain the IceCube PeV neutrino spectrum and DM relic abundance.

It is worth discussing the estimate~\eqref{estimate} in connection with other cosmological bounds on $\kappa$ from recent literature\footnote{Notice that the estimates in~\cite{CosmKan3,CosmKan4} are exhibited in terms of the re-scaled Kaniadakis parameter $\beta=\kappa\frac{M_p^2}{H_0^2}$, where $H_0$ is the present Hubble rate.} (see Tab.~\ref{Tab1}). While being lower than the value $\kappa\simeq0.2$ needed to fit the cosmic rays spectrum~\cite{Kania1}, the obtained $\kappa$ is appreciably non-vanishing if compared, for example, with constraints from Baryon Acoustic Oscillations~\cite{CosmKan3}, cosmological constant and Type Ia Supernova measurements~\cite{CosmKan3}, Hubble, strong lensing systems and HII galaxies data~\cite{CosmKan4}, and 
${}^7 Li$-abundance observations~\cite{CosmKan5}. This suggests that, in principle, the IceCube PeV neutrinos could be more sensitive to the effects of the Kaniadakis entropy~\eqref{KenBH} than other systems/cosmic scenarios, providing a valuable playground to test  Kaniadakis prescription in perspective. 

\begin{table}[t]
  \centering
    \begin{tabular}{|c| c c|}
    \hline
\, Estimate ($|\kappa|$)\,\, &\, Physical framework \, &\,  Ref.\, \\
        \hline
        \,$6\times10^{-125}$\, & 
             Baryon Acoustic Oscillation (BAO) & \cite{CosmKan3} \\[1.5mm]
           \hline
           \,$3\times10^{-125}$\, & 
             CC+SNIa+BAO & \cite{CosmKan3} \\[1.5mm]
           \hline
           \,$1.2\times10^{-124}$\, & Cosmological constant (CC) & \cite{CosmKan3} \\[1.5mm]
           \hline
            \,$1.3\times10^{-124}$\, & 
            Type Ia supernova (SNIa) & \cite{CosmKan3} \\[1.5mm]
            \hline
            \,$3.6\times10^{-123}$\, & 
             Hubble data & \cite{CosmKan4} \\[1.5mm]
           \hline
              \,$4.4\times10^{-123}$\, & 
             Strong lensing systems & \cite{CosmKan4} \\[1.5mm]
           \hline
              \,$3.7\times10^{-123}$\, & 
             HII galaxies & \cite{CosmKan4} \\[1.5mm]
           \hline
             \,$8.1\times10^{-84}$\, & ${}^7 Li$-abundance
              & \cite{CosmKan5} \\[1.5mm]
           \hline
    \end{tabular}
  \caption{Some bounds on Kaniadakis entropic parameter from Cosmology and Astroparticle physics.}
  \label{Tab1}
\end{table}
 
 Although not contemplated in the original 
Kaniadakis formalism, the gap between 
our result and other cosmological bounds on $\kappa$ could be explained  by allowing the entropic parameter to be running. This assumption can be understood in the following picture: in the same way as the energy content (that is, the matter degrees of the freedom)
of the Universe is described by a dynamic fluid evolving from an initially relativistic to a semi- or non-relativistic system as the temperature cools down, we can think of the holographic entropy (i.e. the horizon degrees of freedom) as undergoing a transition from a relativistic (Kaniadakis-type, $|\kappa|>0$) to a classical (Boltzmann-Gibbs-type, $\kappa=0$) description for decreasing redshift. In this framework, the departure~\eqref{KenBH} from the classical entropy would be quantified by a decreasing function of the time (or, equivalently, by an increasing function of the energy scale). This dynamical behavior would also be necessary to satisfy the requirement that $Z_\kappa(T)$ can in principle depart from unity at the pre-BBN epoch, where we still do not have direct constraints by cosmological observations, but it must recover GR (i.e. $Z_\kappa(T)=1$) in the late stages of the Universe evolution for phenomenological consistency (see also the discussion in the Introduction). 

We recall that a similar scenario with a varying deformation entropic parameter has been conjectured in~\cite{LucBlas,Run1,Run2} in the context of non-extensive Tsallis entropy and in~\cite{DiGennaro} for the case of Barrow entropy. In particular, in~\cite{Run1} it is observed that
the renormalization of a quantum theory entails a scale-dependence of the degrees of freedom. In the standard theory of fields, massive modes decouple and the
degrees of freedom decrease in the low energy regime. On the other hand, in gravity theory the situation is more cumbersome, as the degrees of freedom could increase if the space-time fluctuations become large in the ultraviolet regime, while they decrease if gravity is topological, which may be compatible with holography. Either way,  
presuming that a deformation of the standard entropy-area law 
is needed, it would be reasonable to 
assume a dynamic deformation parameter to account for 
these features both at high-energy (inflation) and
low-energy (late-time Universe) scale.
Clearly, more work to consolidate this picture is required, especially in view
of formulating a new relativistic thermodynamics that 
incorporates a running non-extensive entropic parameter
in a self-consistent way.

\section{Conclusion and Discussion}
\label{Disc}

It is a fact that the standard Boltzmann-Gibbs
theory cannot be applied to systems where the partition function diverges, and (large-scale) gravitational systems are known to belong to this class. In these footsteps, recent works proposed a generalization of the holographic dark energy scenario and the cosmological Friedmann equations equipped with the Kaniadakis entropy, which is a one-parameter deformation 
of Boltzmann-Gibbs entropy incorporating special relativity. Motivated by these insights, in the present work we addressed the observed discrepancy between 
the present bound on the Dark
Matter relic abundance and the IceCube high-energy neutrino data in Kaniadakis entropy-based Cosmology. Our strategy was to keep the canonical (4-dimensional) Yukawa-like coupling unchanged, while modifying the description of the Universe evolution by using the $\kappa$-deformed entropy in Eq.~\eqref{KE} (or, equivalently, Eq.~\eqref{KenBH}). By resorting to the generalized Friedmann equations~\eqref{FFE}-\eqref{SFE} and solving the evolution equation of DM particles, we proved that the IceCube neutrino tension can be alleviated in this framework, provided one properly constrains the scaling exponent $\kappa$. This is 
line with other results in recent literature, which show that Kaniadakis
entropy works better than the classical Boltzmann-Gibbs one for a vast class of relativistic and/or complex systems, such as cosmic rays, plasma, open stellar clusters, nuclear collisions processes, etc. 
Since PeV neutrinos fall within this class of systems, the use of a relativistically motivated statistics appears natural and all the more necessary.

Further aspects remain to be investigated:
first, our analysis was performed in the approximation of small departure from the Boltzmann-Gibbs statistics. 
Although this assumption does not undermine the conceptual basis of our study -- since $\kappa\ll1$ is the expected scenario -- 
a more reliable estimation of Kaniadakis parameter should be inferred by exact calculations. This is also requested by the fact that relativistic symmetries are exactly preserved only by the full Kaniadakis entropy.
Due to the peculiar form of Eq.~\eqref{KenBH}, such a task involves more computational effort, which will be conducted in a future extension of this work. 

As additional perspectives, it would be suggestive to compare our approach (and possibly find a connection) with other studies that adopt a different modus operandi to explain
the IceCube PeV neutrino spectrum. 
For instance, in~\cite{Murase} and~\cite{Sahu} 
exotic types of interactions are used.
In particular, in~\cite{Murase} the authors take into account secret interactions of neutrinos with the cosmic background, while in~\cite{Sahu} photohadronic coupling of the Fermi accelerated high energy protons are considered with the synchrotron background photons in the nuclear region of high energy blazars and Active Galactic Nuclei. Finally, a challenging goal is to further explore the possibility to allow for a running $\kappa$. In this sense, it could be helpful to search for
signatures of Kaniadakis entropy in the very early Universe, where the effects of a potential departure from Boltzmann-Gibbs entropy might be amplified. Preliminary clues can be offered by
the study of imprints of the inflationary tensor perturbations~\cite{InfKan} propagated during the hypothetical
Kaniadakis cosmic era in experiments on primordial gravitational waves.
These lines of research are under active investigation and will be presented elsewhere. 

\medskip

\noindent\textbf{Data Availability Statement} All data that have been used in our analysis have already been freely released and have been published by the corresponding research teams. In our text we properly give all necessary References to these works, and hence no further data deposit is needed.

\acknowledgements
GGL acknowledges the Spanish ``Ministerio de Universidades''
for the awarded Maria Zambrano fellowship and funding received
from the European Union - NextGenerationEU.
He is also grateful for participation to the  LISA Cosmology Working group.
GL thanks MUR and INFN for support. 
GL and GGL acknowledge the participation to the COST
Action CA18108  ``Quantum Gravity Phenomenology in the Multimessenger Approach''.

\end{document}